\documentclass[graybox]{svmult}

\usepackage{mathptmx}       
\usepackage{helvet}         
\usepackage{courier}        
\usepackage{type1cm}        
\usepackage{makeidx}         
\usepackage{graphicx}        
\usepackage{multicol}        
\usepackage[bottom]{footmisc}

\makeindex             

\begin{document}

\title*{New ground-based observational methods and instrumentation for asteroseismology}
\author{Pedro J. Amado}
\institute{Pedro J. Amado \at Instituto de Astrof\'{i}sica de Andaluc\'{i}a, Glorieta de la Astronom\'{i}a s/n, \email{pja@iaa.es}}
%
%
\maketitle

\abstract*{Space instrumentation like SOHO, MOST, CoRoT \& Kepler has been and is being built to attain very high precision data to be used for asteroseismic analysis. Nonetheless, there is a very strong need for providing additional information, especially on mode identification. With this contribution I will review the efforts been put on new ground-based instrumentation and the methodology that can be used to achieve this aim.}

\abstract{Space instrumentation like SOHO, MOST, CoRoT \& Kepler has been and is being built to attain very high precision data to be used for asteroseismic analysis. Nonetheless, there is a very strong need for providing additional information, especially on mode identification. With this contribution I will review the efforts been put on new ground-based instrumentation and the methodology that can be used to achieve this aim.}

\section{Introduction}
\label{sec:1}


Topics like rotation, convection, diffusion and settling of heavy elements vs. mixing, magnetic fields and others are a challenge for both stellar structure and evolution and for asteroseismology. These issues, not being fully understood, limit our ability to determine accurate stellar ages, global parameters and the details of the internal structure of the stars.

To help solving the aforementioned challenges, asteroseismology needs the observational input from ground-based instrumentation. However, for asteroseismology to work, the observational inputs need to be as accurate as possible, so that a comparison and refinement of models can be achieved. This observational input can be classified in three groups: 1) {\bf Physical parameters of pulsating stars}, for which ground-based instrumentation has provided a wealth of information for all type of stars in the HR diagram; 2) {\bf pulsation frequencies}, whose quality depends on the ability of the data time series to resolve all the frequencies and determine them with enough precision; and 3) {\bf mode identification}, which requires the comparison between the oscillation theory applied to the stellar atmosphere and the asteroseismic observables like frequencies, their amplitudes and phases or line profile variations (LPVs). Mode identification is probably the most important input asteroseismology needs.

\section{Mode identification}
\label{sec:2}

The approaches followed to tackle this problem can be grouped into three different categories: 1) {\bf Frequency patterns:} in the asymptotic regime, the large and small separation in solar-like oscillations or the use of the Frequency Ratio Method \cite{Moya} for $\gamma$~Doradus stars help reducing the values of the degree and radial order of the modes (see Reed et al., these proceedings). Also, for $\delta$~Scuti stars, quasi-periodicities are been found in a handful of objects which are predicted by the theory, but these periodicities do not necessarily provide mode identification, although it can be used as an additional asteroseismic observable (see Garc\'{\i}a-Hern\'andez et al. and Su\'arez et al., these proceedings); 2) {\bf direct fitting} of observed and computed frequencies considering all quantum number values is difficult, especially when small or very large number of frequencies are observed, or when they do not show frequency patterns; and 3) {\bf mode identification of a single or a very few modes}, especially of the main pulsating modes in the star, can imply a large reduction of models fitting the observations. This last approach is the one that can benefit most from ground-based observations.


To carry out the identification of the single modes there are two types of data which need to be taken as time series: 1) {\bf multicolour photometry}, which provides information from the different amplitude and phase behaviour of the modes at different wavelengths bands, thus providing information of the degree $\ell$ of the mode (\cite{Garrido}; see also Pamyatnykh  et al. in these proceedings); and 2) {\bf high-resolution spectroscopy} allows us to study LPVs with a set of different methods.

To study LPVs time series from these spectroscopic data, two approaches can be followed: 1) {\bf to careful select a number of lines}, which depends on the spectral type of the star \cite{DeRidder,Zima}, they must be unblended, deep, thermally broadened (avoid hydrogen lines), etc; and 2) {\bf averaging all lines in the spectrum} \cite{Uytterhoeven} providing a high SNR mean line profile by using optimization techniques like the LSD \cite{Donati}, cross-correlation, etc. The drawback of this latter method is that it assumes that all the lines have the same intrinsic profile and that they form in the same region of the atmosphere, which may not be the case.


\section{Past and current instrumentation for asteroseismology}
\label{sec:4}

The various types of data and techniques mentioned above have been used for a long time. For instance, the discovery of periodic variable stars begun back in the seventeenth century and Struve was already detecting LPVs in the spectrum of the star $\delta$~Sct in 1953 \cite{Struve}. The list of instruments providing these data is enormous and, therefore, just a few milestones that have occurred in the past few years will be briefly mention: 1) {\bf Small and medium-sized telescopes} all over the world have been collecting optical photometry or spectroscopy, including APTs, in a variety of photometric indices (Johnson, Stroemgren, Geneva, etc) with photoelectric photometers and with CCDs detectors; 2) {\bf coordinated networks} set up by both the Solar (GONG, BiSON, IRIS, TON) and the Stellar (WET, DSN, STEPHI or multi-site campaigns with different telescopes) communities have been used to obtain observations all around the world. They have improved the precision in the determination of the frequencies, avoiding strong aliases problems with much cleaner spectral windows; 3) {\bf development of high-accuracy RV spectrographs}, which have allowed in the past few years the detection and eventually the confirmation of the presence of solar-like oscillations in stars other than the Sun: HARPS, UVES, CORALIE, SARG, UCLES, SOPHIE, FIES, HERMES, HERCULES, etc. They have contributed to the refinement of the stars' global parameters and provided mode identification. However, there have been very few coordinated multi-site campaigns in the last years, two of which are the spectroscopic follow up for CoRoT (\cite{Uytterhoeven09}; Poretti et al., these proceedings) or the coordinated campaign for Procyon (11 spectrographs; \cite{Arentoft}); 4) {\bf Large Surveys}: OGLE, MACHO, RATS, irsf/sirius, VVV\footnote{http://mwm.astro.puc.cl/mw/index.php/Main\_Page}, etc; finally 5) {\bf Interferometry} is a very young field with a bright future which can contribute to asteroseismic studies by providing accurate radii for stars and perhaps much more (see contributions in these proceedings on VLTI, SUSI, KI, PTI, NPOI, CHARA, COAST)\footnote{http://en.wikipedia.org/wiki/List\_of\_astronomical\_interferometers\_at\_visible\_and\_infrared\_wavelengths}.

All these instruments and projects have contributed to the three basic inputs needed for asteroseismology named in Sect.~\ref{sec:2}. However, there are two main areas that have not been fully exploited by current instrumentation for asteroseismology, viz., stellar networks for high-resolution spectroscopy or from the single high-duty-cycle site of Antarctica, and the NIR wavelength observations (ultraviolet is another unexplored wavelength region but it needs space instrumentation).

\noindent{\bf Long, high-duty-cycle RV/LPVs time series:}
The solution for obtaining these high-quality time series is either to implement a worldwide network of hires, high-accuracy spectrographs or to implement a single one in a site like Antarctica. An example of the first is SONG (Stellar Observations Network Group; \cite{Grundahl}) which will set up a network of eight alt-azimuthal 1-m telescopes providing 1m/s precision radial velocities, and photometry of crowded fields with a dual-channel lucky imaging camera with up to 6 filters. An example of the second is SIAMOIS (Seismic Interferometer Aiming to Measure Oscillations in the Interior of Stars; \cite{Mosser}), which is a project to achieve high precision Doppler velocities from Dome C that could carry out a scientific programme complementary to CoRoT and Kepler. The core of the instrument, providing the necessary sensitivity and stability, is a Fourier tachometer similar to the helioseismic network GONG, fully automated and without moving parts. The high duty cycle (up to 90\%) is comparable to space-borne observations.



\noindent{\bf Near-Infrared wavelength observations for asteroseismology}
A list of reasons for the interest of observing pulsating stars in the near infrared (NIR) is: 1) it will provide a new wavelength region from which to obtain stellar physical parameters, especially for red giants, which show a sufficient number of lines and molecular bands; 2) the flux peak moves toward NIR wavelengths for cooler objects, providing thus higher SNR data; 3) red giants show larger amplitudes in the NIR than at VIS wavelengths; 4) NIR observations probe different parts of the line forming region in stellar atmospheres with respect to the VIS wavelengths. Therefore, they can improve mode identification.

The main reason to explain why there have been no systematic observations of pulsating stars in the NIR wavelength regions is because of the poor precision achieved with NIR detectors. This is no longer the case, as absolute photometric accuracies better than 10 mmag can be achieved in the $JHK$ bands \cite{Sollima,Minniti}. Other reasons for using NIR photometry are: 1) NIR commercial detectors allow two-channel cameras to simultaneously observe in the VIS and in the NIR in small telescopes; 2) red and NIR wavelengths provide increased discrimination of pulsating modes \cite{BalonaEvers}; 3) the $Y$ band is accessible without cryogenics, $JH$ bands need cooling though. Reasons for using NIR spectroscopy are: 1) there is less crowding of lines, implying less blending in high-velocity rotators which in turns means that LPVs are more clearly detected; 2) NIR observations are expected to provide increased discrimination of pulsating modes in spectroscopy through differences in amplitudes and phases in the NIR when compared with the VIS \cite{Amadoa,Amadob}.  



To exploit the increased mode discrimination of red and NIR wavelengths, photometric instruments should provide simultaneous VIS and NIR light curves. A few examples of current and future instrumentation are: REM@LaSilla is a 60 cm diameter fast reacting telescope with two instruments REMIR, an infrared imaging camera, and ROSS, a visible imager and slitless spectrograph which can work simultaneously. GROND at MPI/ESO2.2m@LaSilla is an imaging instrument to take observations simultaneously in 4 visible and 3 NIR channels. Bootes IR@OSN in Granada is another 60 cm diameter fast reacting telescope with a CCD detector and soon will be equipped with an IR camera working in the NIR bands. MAIA@MERCATOR will provide fast photometry in three simultaneous wavebands in the SDSS system u', g' r'+i'. PANIC@2.2m of CAHA is a panoramic camera in the NIR with a mosaic of 4 Hawaii-2RG IR detectors which will provide accurate NIR light curves of variable stars in a very wide FoV.

However, the azimuthal order, $m$, cannot be determined by photometric means alone. Each mode of pulsation introduces a distinctive velocity field on the surface of the star. This is observed in the stellar spectrum as a periodic variation in the line profiles with high resolution spectroscopy. And it is here where NIR spectroscopy can make a real change. However, the existing high resolution NIR spectrographs do not have cross-dispersion, covering therefore a very small range in wavelength, or do not have sufficient resolution for RV and LPV work. Some future projects will deliver instruments providing high resolution and large wavelength coverage. CARMENES (Calar Alto high-Resolution search for M dwarfs with Exo-earths with Near-infrared and optical Echelle Spectrographs; \cite{Quirrenbach}) is one of these instruments. It is a next-generation instrument to be built for the 3.5m telescope at the Calar Alto Observatory by a consortium of eleven Spanish and German institutions. Conducting a five-year exoplanet survey targeting ~300 M dwarfs with the completed instrument is an integral part of the project. The CARMENES instrument consists of two separate echelle spectrographs covering the wavelength range from 550 to 1700 nm at a spectral resolution of R = 82\,000, fed by fibers from the Cassegrain focus of the telescope. The spectrographs are housed in vacuum tanks providing the temperature-stabilized environments necessary to enable a 1~m/s radial velocity precision employing a simultaneous calibration with an emission-line lamp. For late-M spectral types, the wavelength range around 1000 nm ($Y$ band) is the most important wavelength region for radial velocity work. The instrument concept is based in a design with two spectrographs, one equipped with a CCD for the range 550-1050 nm, and one with HgCdTe detectors for the range from 900-1700 nm.


CARMENES asterosesimology science cases are: 1) mode identification in the NIR, 2) Pulsations in M dwarfs (see Rodr\'{\i}guez-L\'opez et al. and Moya et al., these contributions), 3) Pulsating stars with planets (see Moya et al. and M. Oshagh et al., these proceedings), 4) Cepheids (Nardetto et al., these proceedings), 5) Solar-like oscillations in red giants (De Ridder et al., Hekker et al., these proceedings).


\section{Future of ground-based instruments useful for asteroseismology. Conclusions and final remarks}
\label{sec:7}

Unfortunately, there is not much to list as future ground-based instrumentation for asteroseismology due to the fact that, either this author is missing part of the instrumental development for asteroseismology for the near future, or there is not much effort/resources being put into this area. Synergies with exoplanet research is still providing very interesting projects but probably little time will be awarded to pure asteroseismic research. These projects involve the development of visible spectrographs like ESPRESSO@VLT, CODEX@E-ELT and NIR spectrographs like CARMENES, SpiRou and UPF. Interferometric instrument is another area of development useful for asteroseismolgy research.

Ground-based instrumentation is essential for asteroseismology. The determination of accurate physical parameters, especially for science preparation of expensive space missions, need more resources for a homogeneous determination and for cross-calibrations with different methods: photometry, spectroscopy, interferometry, etc. Space photometry provides accurate pulsation frequencies. However, solar-like oscillations can benefit from VIS and NIR hires spectroscopy, especially for red giants. There are new areas of development still to be exploited: spectroscopic networks in the VIS, and NIR spectrographs which can lead to the development of VIS+NIR spectroscopic networks.

Space missions like CoRoT, Kepler and maybe PLATO are and will continue to revolutionize the field of asteroseismology. Nonetheless, a few final questions put forward by the author for discussions are the following: are the resources invested in space instrumentation followed by sufficient resources for ground-based instrumentation? And for new theoretical developments (human resources)? What would be the results for asteroseismology if similar resources to those allocated for space missions were used in the aforementioned items? Could this ground-based instrumentation exploit methods which have not yet been fully tested? Is the access of the asteroseismic community to very large telescope sufficient? Exoplanet research has driven asteroseismology research through the synergies in techniques and instrumentation, but how much can we exploit those synergies? and, is asteroseismology research being held back by the limitations impose by exoplanet research? Is the asteroseismic community strong enough to drive the development of their own specific instrumental project roadmap? Answer to these questions might be provided in the near future.

\begin{acknowledgement}
The author acknowledges financial support from projects AYA2009-08481-E and AYA2010-14840 from the Spanish Ministry of Sciene and Innovation.
\end{acknowledgement}

\end{document}